\documentstyle[epsfig]{aipproc}

\begin{document}
\title{Observations of Mkn 421 with the CELESTE Experiment}

\author{J. Holder$^*$ for the CELESTE collaboration$^{\dagger}$}
\address{$^*$ LAL, Universit\'{e} Paris Sud, Orsay 91405, France\\
$^{\dagger}$CESR, Toulouse 31029, France,\\
CEN de Bordeaux-Gradignan 33175, France,\\
PPC, Coll\`{e}ge de France, Paris 75231, France,\\
LPNHE, Ecole Polytechnique, Palaiseau 91128, France,\\
GPF, Universit\'{e} de Perpignan 66000, France,\\
Charles University, Prague 18000, Czech Republic,\\
Joint Laboratory of Optics, Olomouc, Czech Republic.\\
}

\maketitle

\begin{abstract}
The CELESTE experiment uses the heliostats of an old solar farm in the French Pyrenees to detect gamma ray air showers by the atmospheric Cerenkov technique.
Observations of the TeV blazar Markarian 421 have been made with the fully instrumented CELESTE experiment since December 1999.
The detection of gamma ray emission from this source at energies greater than 50 GeV is presented here.
A comparison is made with the light curve from the CAT experiment at the same site which shows correlation between the observed gamma ray fluxes and the detection of short duration flaring episodes.
\end{abstract}

\section*{Introduction}

CELESTE has been in stable operation with a 40 heliostat array since October 1999. Our main targets during this initial period have been the three most reliably detected sources in the TeV sky; the Crab nebula, Markarian 421 and Markarian 501. Our detection of the Crab nebula is described elsewhere in these proceedings \cite{crabceleste}. This report is concerned with the observations of Mkn 421.

Mkn 421 is the closest X-ray selected BL Lac object (z=0.030) and was the first extragalactic object detected at TeV energies \cite{punch}.
The emission from this source is extremely variable over timescales as short as 15 minutes \cite{gaidos} and multiwavelength observations have shown that the TeV variability is correlated with the X-ray emission \cite{buckley}.
Observations in the 50~GeV region, not possible until now, will help to improve our understanding of the emission processes at work.


The CELESTE experiment is described in detail in the experimental proposal \cite{proposal} and in \cite{NIM,berry}.
Briefly, CELESTE uses forty, $54\ m^{2}$ heliostats of a former solar electrical plant at the Themis site in the French Pyr\'en\'ees (N. $42.50^{\circ}$, E. $1.97^{\circ}$, altitude 1650\ {m}).
The light from the seperate heliostats is resolved using secondary optics at the top of a 100~m tall tower.
At the secondary mirror focus is one PMT for each heliostat. A solid Winston cone at the face of each PMT defines the field of view of each tube to be 10~mrad.

The trigger is constructed by summing eight PMT signals in five groups and sending each sum to a discriminator with a threshold set to 4.5 photo-electrons (PE) per heliostat.
A logic coincidence of at least three of the five groups triggers the experiment.
Each PMT signal is sent to a 1~GHz flash ADC circuit which digitizes the signal.
When a trigger occurs, digitization stops and a window of 100~ns centered
at the nominal \v{C}erenkov pulse arrival time is read out.

\section*{Analysis}

CELESTE, with its large mirror area per PMT, is very susceptible to systematic effects in the data caused by differences in the night sky background light between the ON source observation and OFF source control data.
In the case of Mkn 421, a star of magnitude 6.1 located at the source position causes a difference of $\sim$15\% in the PMT anode currents between the two fields.
To compensate, we have adapted the software padding method proposed by Cawley \cite{cawley} for use with our FADC data \cite{crab}.
We also apply a software trigger by using the FADC data to reconstruct the five trigger group sums and then demanding:$\ge$4 groups $>$5.0\ PE per heliostat.
This increases the energy threshold, but provides us with comparable background data in the ON and OFF source fields. We also require $>10$~\v{C}erenkov peaks with an amplitude $>25$~digital counts to ensure the shower parameters are  measurable.

The rest of the analysis is concerned with improving the gamma ray signal to hadron background ratio using two parameters. The first, $\sigma_{grp}$, is a measurement of the homogeneity of the light distribution between the trigger groups and should be the most efficient background rejection cut according to our simulations, rejecting 80\% of the remaining hadrons for only 30\% of the gammas.
The second parameter, $\theta$, is the shower axis angle relative to the pointing direction reconstructed using two points; the barycentre of the light distribution at ground level and the position of the shower core relative to the pointing direction at the altitude of maximum emission deduced from a sphere fit to the \v{C}erenkov pulse arrival times at the heliostats.

\section*{Observations and Results}

Mkn 421 was observed between November 1999 and March 2000.
After rejecting those observations with bad weather or equipment problems, the dataset consists of 80 ON/OFF pairs with an ON source exposure of 22.5 hours.

\begin{figure}[tr] 
\centerline{\epsfig{file=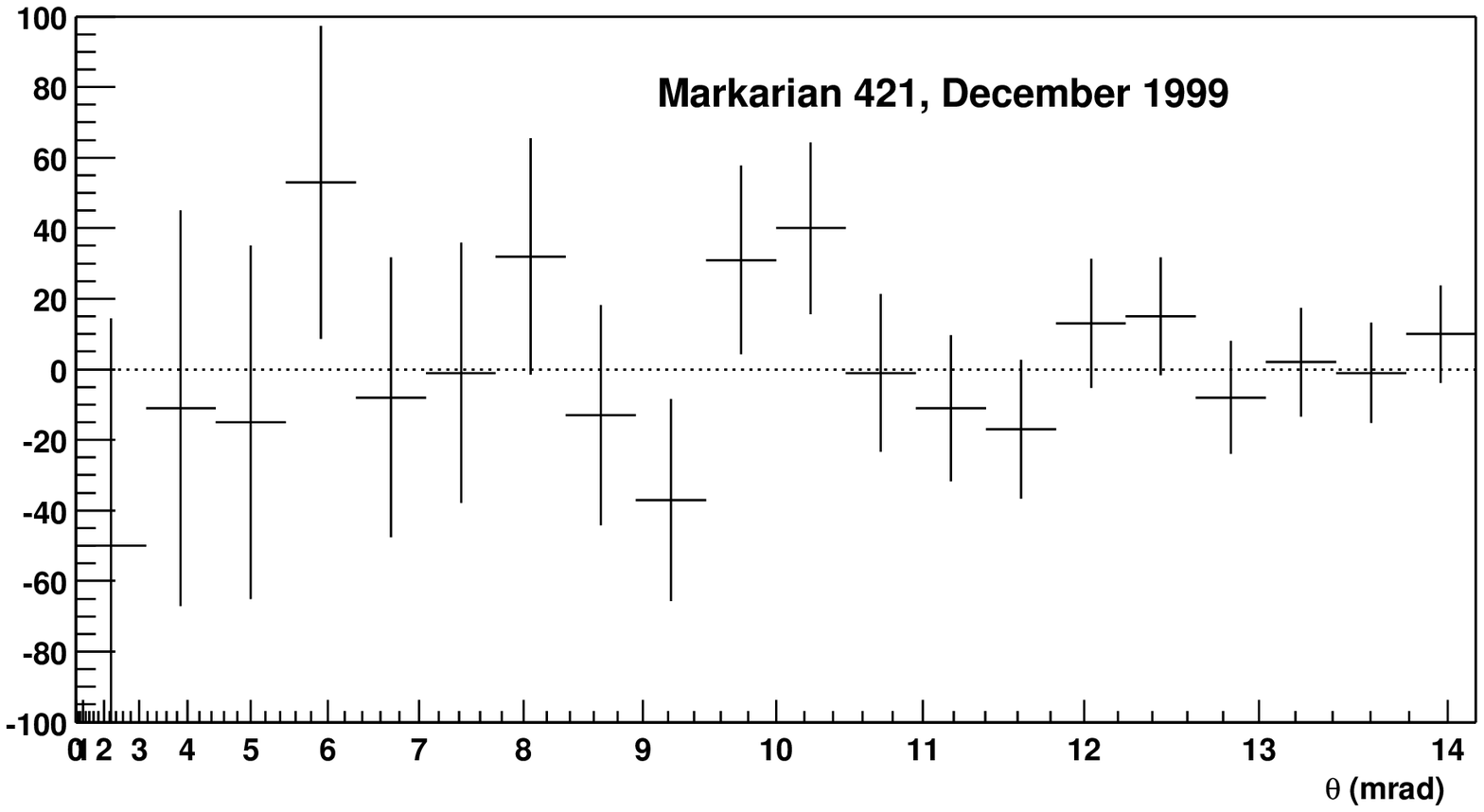,height=1.7in,width=3.0in}\epsfig{file=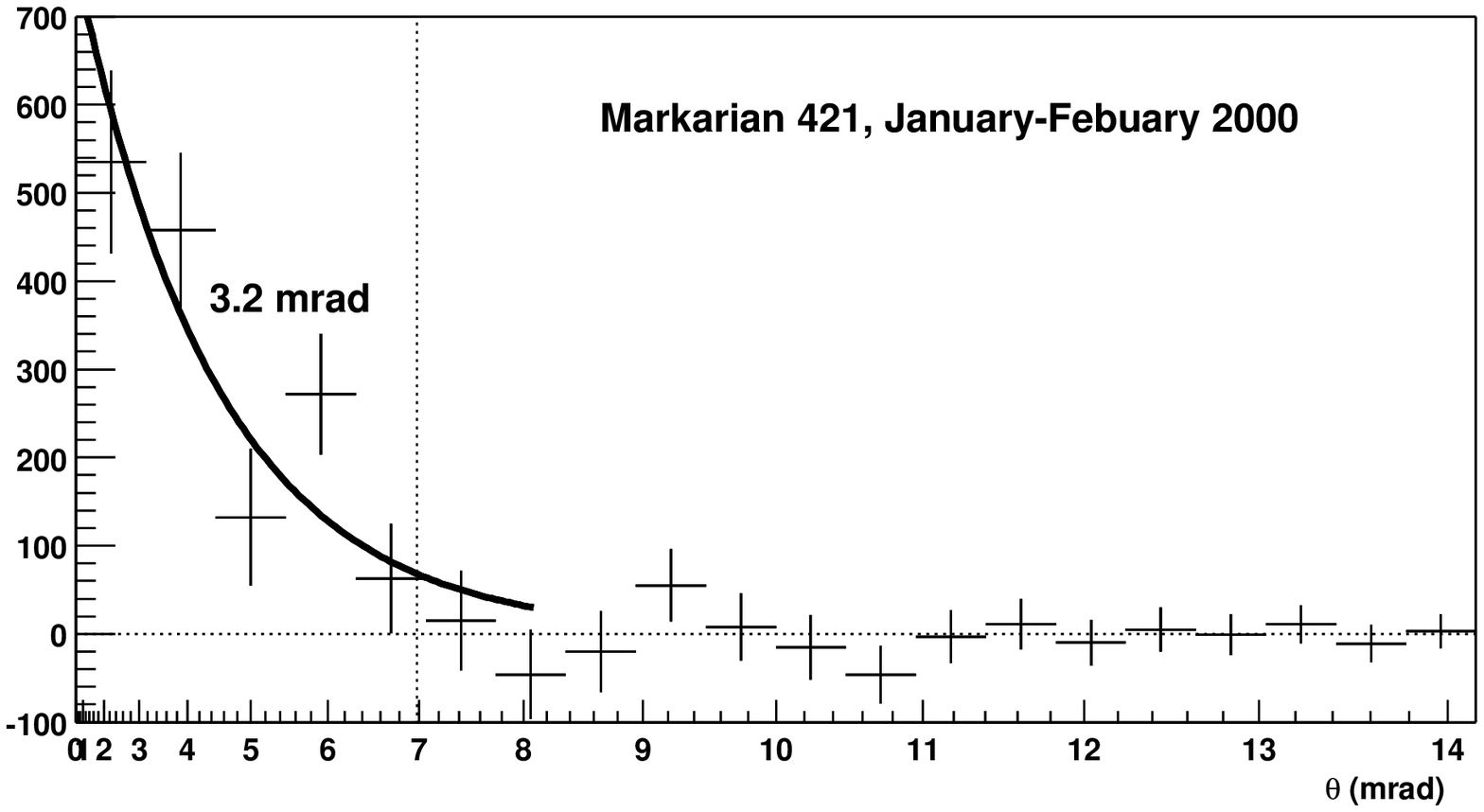,height=1.7in,width=3.0in}}
\vspace{10pt}
\caption{The difference in the distribution of the shower axis angle $\theta$ between ON and OFF source observations of Mkn 421. {\it Left:} For the December 1999 quiet period. {\it Right:} For the January/February 2000 active period.}
\label{theta}
\end{figure}

The CAT imaging telescope \cite{barrau}, situated on the same site as CELESTE, made observations of Mkn 421 during the same period.
In December 1999, no evidence for a signal was reported by CAT.
The left-hand plot of Fig.~\ref{theta} shows the distribution of the difference in the shower orientation parameter, $\theta$, between the ON and OFF source observations as measured by CELESTE during December. 
No evidence for a signal at low $\theta$ is seen (for 1.5 hours of ON source exposure).
Table~\ref{tquiet} shows the number of events remaining at each stage of the analysis.

\begin{table}[b!]
\caption{\label{tquiet} The number of events, excess, significance and signal to background ratio (S/B) after each cut for the December quiet period.}
\begin{tabular}{l|ccccc}
 & ON & OFF & Excess & Significance & S/B \cr
\hline
Raw                         &  $104\,443$ & $10\,3991$ & $452$ &  $1.0$\hspace{1em} &  $0.4\%$\hspace{1em} \cr
Software Trigger       &   $55\,294$ &  $55\,236$ & $58$  &  $0.2$\hspace{1em} &  $0.1\%$\hspace{1em} \cr
$N_{\mathrm{peaks}}\geq 10$   &   $51\,943$ &  $51\,924$ & $19$  &  $0.1$\hspace{1em} &  $0.0\%$\hspace{1em} \cr
$\sigma_{\mathrm{grp}}<0.25$ &   $12\,154$ &  $12\,186$ & $-32$ & $-0.2$\hspace{1em} & $-0.3\%$\hspace{1em} \cr
$\theta\leq 7~\mathrm{mrad}$       &    $6\,648$ &   $6\,679$ & $-31$ & $-0.3$\hspace{1em} & $-0.5\%$\hspace{1em} \cr
\end{tabular}
\end{table}
In January/February 2000 CAT reported that Mkn 421 was in an active state. In particular, two bright flares were observed, with emission levels $>$2 Crab.
The right-hand plot of Fig.~\ref{theta} shows the difference distribution in $\theta$ for CELESTE observations during these flares (5.2 hours of ON source exposure).
A strong signal is apparent at low values of $\theta$, as expected for gamma rays. The significance of the excess is 8.1~$\sigma$. 
Table~\ref{tflare} shows the improvement in the signal/background ratio at each stage of the analysis after the initial cleaning cuts. The significance before cuts is not very meaningful as the excess, while undoubtably largely due to gamma rays, is contaminated by systematic effects due to sky noise differences.
\begin{table}[t!]
\caption{\label{tflare} The same as Table~\ref{tquiet} for the January/February active period.}

\begin{tabular}{l|cccccc}
 & ON & OFF & Excess & Significance & S/B  & $\gamma/\mathrm{min}$\cr
\hline
Raw                         &  $369\,424$ & $362\,248$ & $7\,176$ & $8.4$\hspace{1em} & $2.0\%$\hspace{1em} &\cr
Software Trigger       &  $177\,321$ & $174\,273$ & $3\,048$ & $5.1$\hspace{1em} & $1.7\%$\hspace{1em} & 9.8\cr
$N_{\mathrm{peaks}}\geq 10$   &  $167\,385$ & $164\,345$ & $3\,040$ & $5.3$\hspace{1em} & $1.8\%$\hspace{1em} & 9.8\cr
$\sigma_{\mathrm{grp}}<0.25$ &   $29\,049$ &  $27\,536$ & $1\,513$ & $6.4$\hspace{1em} & $5.5\%$\hspace{1em} & 4.9\cr
$\theta\leq 7~mrad$       &   $17\,169$ &  $15\,709$ & $1\,460$ & $\mathbf{8.1}$\hspace{1em} & $9.3\%$\hspace{1em} & 4.7\cr
\end{tabular}
\end{table}

\begin{figure}[t!] 
\centerline{\epsfig{file=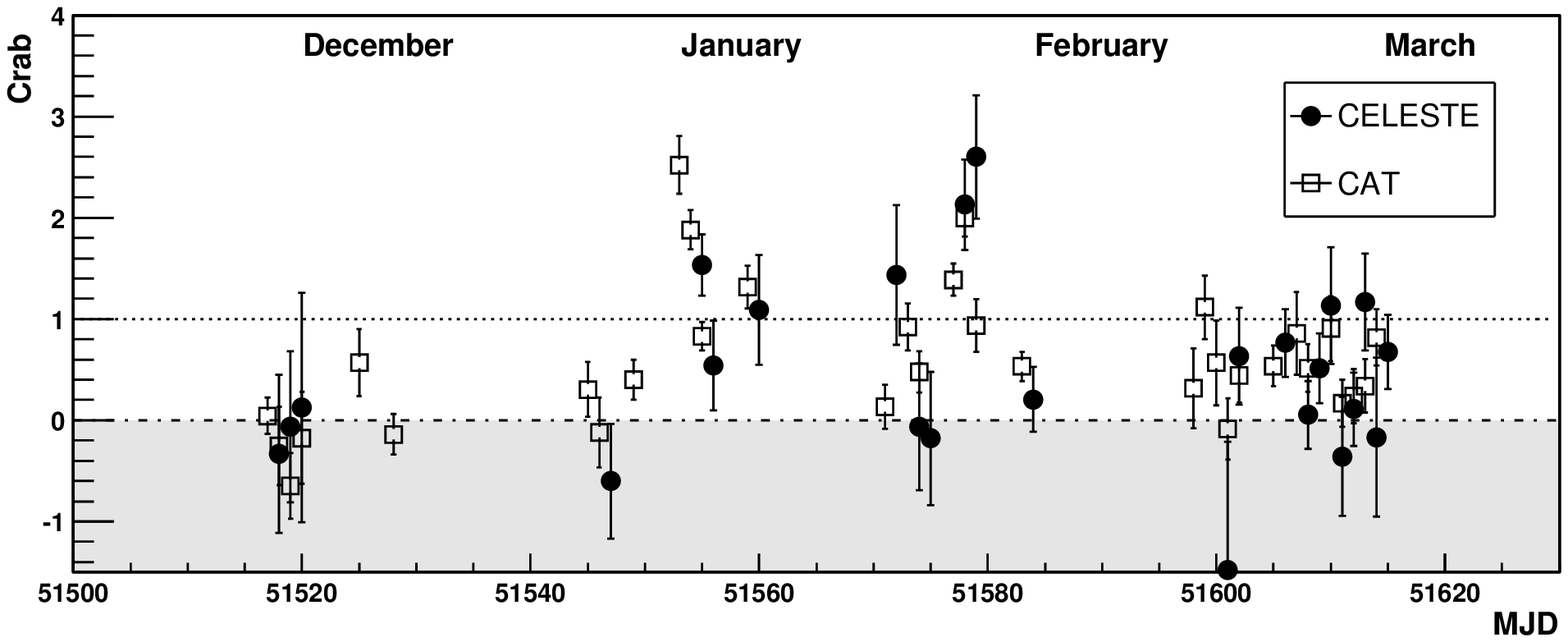,height=1.7in,width=4in}\epsfig{file=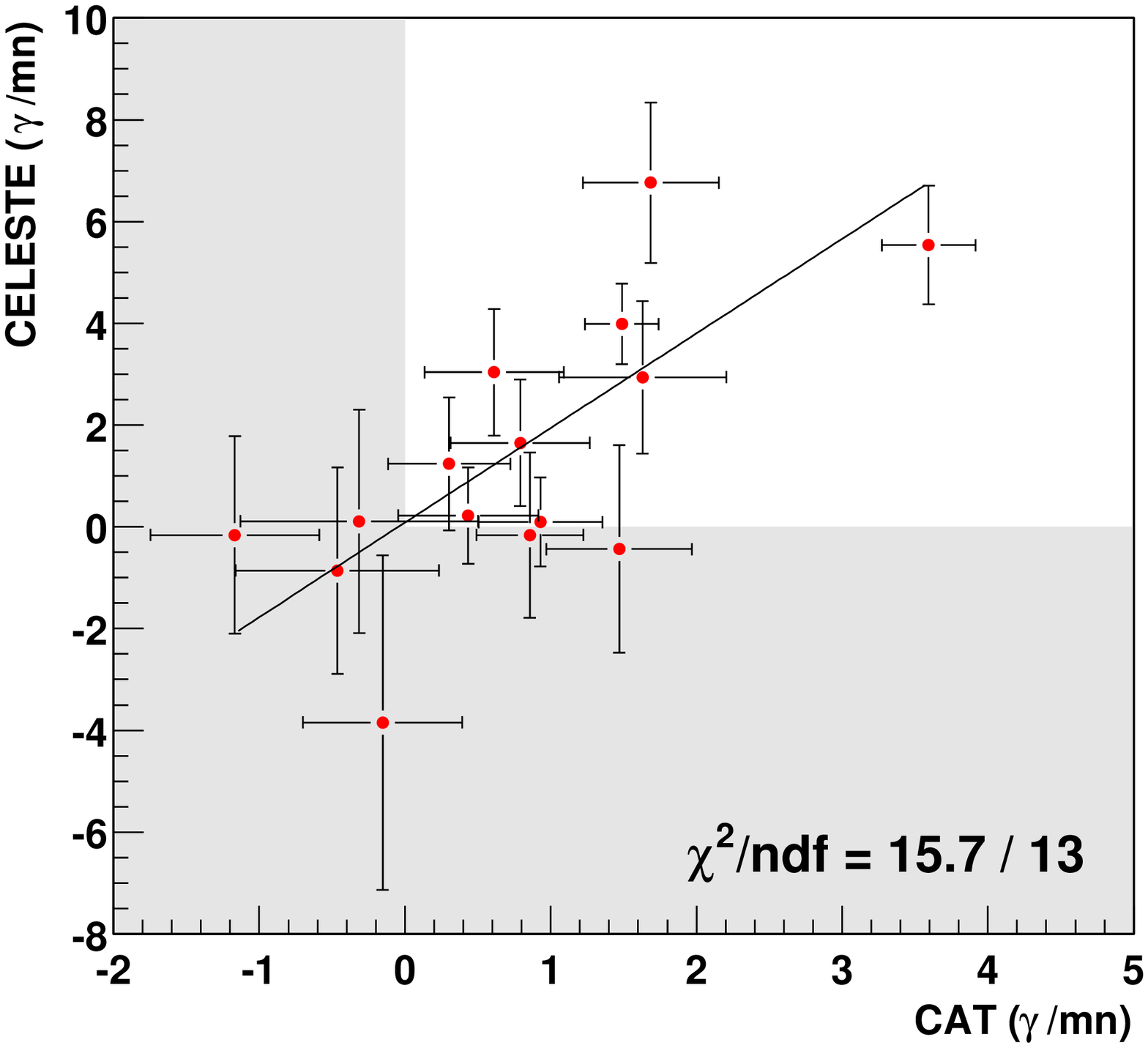,height=1.7in,width=2.in}}
\vspace{10pt}
\caption{The CAT/CELESTE night by night light curve for Mkn 421 and the correlation between the experiments for measurements taken on the same nights.}
\label{lightcurve}
\end{figure}

The whole Mkn 421 dataset has been analysed and Fig.~\ref{lightcurve} shows the lightcurve as observed by both CAT and CELESTE on a night by night basis.
Emission levels up to $\sim3$~Crab ($\sim7~\gamma~\mathrm{min}^{-1}$) were observed, detectable by CELESTE in a single night. 
Also shown in the figure is a plot of the gamma ray flux measured by CELESTE against that measured by CAT showing reasonably good correlation.

\section*{Conclusions}

The TeV blazar Mkn 421 has been detected by CELESTE for the first time at an energy of $\sim50$~GeV.
Evidence for night by night variablility has been observed, and the variablity measurements by CELESTE are correlated with those of CAT.
Importantly, a non-detection of the source in a quiescent state, followed by a very significant detection in a flaring state demonstrates that the CELESTE analysis is working, and that we have been able to correct for systematic effects due to night sky light differences between the ON and OFF source fields.


\begin{references}

\bibitem{crabceleste}de Naurois M., et al., {\it These Proceedings}.
\bibitem{punch}Punch M., et al., {\it Nature}\ {\bf 358}, 477 (1992).
\bibitem{gaidos}Gaidos, J.A. et al., {\it Nature}\ {\bf 383}, 319 (1996).
\bibitem{buckley}Buckley, J.H. et al., {\it ApJ}\ {\bf 472}, L319 (1996).
\bibitem{proposal}{\it CELESTE proposal} (\verb+http://polywww.in2p3.fr/celeste/public/cxp.ps.gz+)
\bibitem{NIM} Reposeur, T., et al., \textit{in preparation}.
\bibitem{berry} Giebels, B., et al., {\it Nucl. Instrum. Meth.}\ {\bf A412}, 329 (1998).
\bibitem{cawley} Cawley, M.F. et al., in {\it Towards a Major Atmospheric \v{C}erenkov Detector 2}, Calgary, 1993, p176. 
\bibitem{crab} de Naurois, M., et al., \textit{in preparation} 
\bibitem{barrau} Barrau, A., et al., {\it Nucl. Instrum. Meth.}\ {\bf A416}, 278 (1998).

\end{references}
\end{document}